\documentclass[12pt]{article}
\usepackage[dvips]{graphicx}
\usepackage{amsmath}
\usepackage{graphicx}
\usepackage{amsfonts}
\usepackage{amssymb}
\usepackage{epsfig}
\usepackage{subfigure}

\newcommand{\be}{\begin{equation}}
\newcommand{\ee}{\end{equation}}
\newcommand{\bea}{\begin{eqnarray}}
\newcommand{\eea}{\end{eqnarray}}
\newcommand{\ba}{\begin{array}}
\newcommand{\ea}{\end{array}}
\newcommand{\beas}{\begin{eqnarray*}}
\newcommand{\eeas}{\end{eqnarray*}}
\newcommand{\bes}{\begin{equation*}}
\newcommand{\ees}{\end{equation*}}

\def\i2           {\mbox{$\frac{i}{2}$}}

\begin{document}
\title{\bf  {\AE}ther Field, Casimir Energy and Stabilization of The Extra Dimension}

\author{ A. Chatrabhuti$^{a}$\thanks{Email:auttakit@sc.chula.ac.th}\hspace{1mm}, P. Patcharamaneepakorn$^{a, b}$\thanks{Email:preeda\_patcharaman@hotmail.com}\hspace{1mm}and P. Wongjun $^{a}$\thanks{Email:pitbaa@gmail.com} \\
$^a$ {\small {\em  Theoretical High-Energy Physics and Cosmology Group, Department of Physics,}}\\
{\small {\em Faculty of Science, Chulalongkorn University, Bangkok
10330, Thailand}}\\
$^b$ {\small {\em  Jawaharlal Nehru University (JNU), New Delhi , India}}\\
}

\maketitle

\begin{abstract}
\noindent In our five-dimensional cosmological model, we investigate the role of a Lorentz violating vector ``{\ae}ther" field on the moduli stabilization mechanism.  We consider the case of a space-like {\ae}ther field on a compact circle with Maxwell-type kinetic term. The Casimir energy of certain combinations of massless and massive bulk fields generates a stabilizing potential for the radius of the compact direction while driving the accelerated expansion in the non-compact directions. It is shown that the {\ae}ther field can reduce the influence of the Casimir force and slow down the oscillation of the radion field.  This property proves crucial to the stability of the extra dimension in the universe where non-relativistic matter is present.  We speculate that this scenario might reveal a hidden connection between the dimensionality of spacetime and the spontaneous breaking of Lorentz symmetry. 
\vspace{5mm}


\end{abstract}

\newpage
\section{Introduction}

Although there have been a lot of progresses on constructing phenomenologically viable models based on theories with extra spatial dimensions, some fundamental questions have not been completely solved.  One of them is the moduli stabilization problem.  The size and shape of compact space described by dynamical moduli fields have to be fixed in order to avoid any conflict with astronomical observations.  In addition to these problems, we also face the challenges from cosmology in explaining the accelerated expansion of the universe. One possible solution for these problems may involve arguments based on anthropic principle. However, the search for an alternative solution is still going on, for example in \cite{Brandenberger}.

Recently, it was suggested that Casimir energy from various field fluctuations in compact extra dimensions could play a crucial role in addressing these significant problems \cite{Ponton, Greene}.  Greene and Levin \cite{Greene} argued  that if the total Casimir energy is properly chosen, then it is possible, at least in the case of vacuum dominated universe, to stabilize the size of the extra dimensions and drive the accelerated expansion of the three non-compacted directions in which the Casimir energy plays the role of dark energy.  The authors in \cite{Burikham} employed the calculation of Casimir energy in the non-trivial space, ${\mathbb M}^{1+3} \times {\mathbb T}^{2}$ and demonstrated that the shape of extra dimensions can also be stabilized by the same mechanism. Interestingly, predictions in this scenario such as radius of extra dimensions and quantum gravity scale in the bulk are in agreement with those from the large extra dimensions or ADD scenario \cite{ADD, antoniadis}.  Hence, the moduli stabilization problem, dark energy problem and the hierarchy problem may possibly be explained in the single unified framework.  However, as it was pointed out in \cite{Greene}, there are some crucial obstructions to realizing a phenomenological viable version of this scenario.  One of them is that the extra dimension fails to stabilize if we include contribution from matter contents.  During the matter dominant epoch, energy density of non-relativistic matter was the dominant contribution in the effective potential of the moduli fields and washed away the minimum of the effective potential.  Although the minimum reappears in the vacuum dominated epoch, the moduli (i.e. radion field) has already passed the dynamical stable fixed point.  This causes the extra dimension to expand and contradicts with our observation.  Thus, it would be interesting to investigate whether this technical problem could be solved.

In this paper, we propose the new stabilization mechanism based on the Casimir energy and the existence of the Lorentz violating ``{\ae}ther" field in the compact direction.  Starting with the simplest model with one extra dimension where the space-like {\ae}ther field lives in the compact circle similar to the model considered in \cite{Rizzo, Carroll_Tam}, we claim that non-vanishing vacuum expectation value (vev) of the {\ae}ther field would affect the dynamical equation of background moduli field.  It can reduce the gradient of the radion's potential and slows down the oscillation frequency.  This ensures stability of extra dimension although there is non-relativistic (dark) matter in the universe.  As in the previous works, the Casimir energy of massless and massive fields embedded in five-dimensional space play a role of dark energy and drive the expansion of non-compact space as expected.  Note that the effect of a time-like {\ae}ther field on slowing down the expansion rate of the universe was pointed out in \cite{Carroll_Lim}. The effects of the {\ae}ther fields on cosmological observable was studied in \cite{Mota}.  The authors in \cite{Obousy_Cas, Obousy_brane} also studied the role of {\ae}ther field on the stability of the extra dimension in the context of braneword scenario but in a different aspect.  

It is important to state that we are aware of the stability issue for space-like {\ae}ther field \cite{Carroll08, Ackerman} which may cause difficulty in the construction of a more realistic model of this scenario.  However the interplay between the {\ae}ther field and the dynamical moduli field in our model may shed some light on the connection between the dimensionality of spacetime and the violation of Lorentz symmetry.  Perhaps nature allows us to observe only the large three-dimensional space that preserves Lorentz symmetry but conceals the Lorentz violating directions in the compact space.

This paper is organized as follows.  In section \ref{aether}, we start by reviewing the {\ae}ther model in 5-dimensional spacetime.  In section \ref{Cosmology}, we derive cosmological equations of motion in 5-dimensional spacetime with {\ae}ther field and write down effective 4-dimensional equations of motion in the radion picture.  Then we review the calculation of Casimir energy and extend to the case involving interaction between {\ae}ther field and bulk fields in section \ref{Casimir}.  In section \ref{stability}, we investigate the role of {\ae}ther field in the context of stabilization of the extra dimension both in the vacuum dominated universe and in the universe with non-relativistic matter.   Finally we summarize our results in section \ref{Conclusions}.

\section{{\AE}ther Field and Its Interactions}\label{aether}

We start by considering a 5-dimensional flat spacetime with coordinates $x^a = (x^{\mu},y)$ where $\mu = 0,\dots,3$ and with mostly plus metric signature.  We assume that the fifth direction is compactified on a circle. Now we consider a toy model in which Lorentz symmetry is spontaneously broken by the {\ae}ther field $u^{a}$ i.e. a vector field with a non-vanishing expectation value.  Most of the {\ae}ther models contain kinetic term that makes their Hamiltonian unbounded from below and their stability is a subtle issue \cite{Carroll08}.  Here we consider the action with Maxwell-type kinetic term \cite{Carroll_Tam}
\begin{eqnarray}
S= \int d^5x \sqrt{-g}\Big( -\frac{1}{4}V_{a b}V^{a b}-\bar{\lambda}(u_a
u^a - v^2)+\sum_i \mathcal{L}_i\Big).
\label{aether_action}
\end{eqnarray}
Here $V_{ab}= \nabla_a u_b - \nabla_b u_a$ has a familiar form to the field strength tensor of electromagnetism. However, the {\ae}ther field $u^a$ is not related to the electromagnetic vector field $A^a$ and its dynamics does not respect $U(1)$ gauge symmetry. In contrast, the second term in the above action enforces the {\ae}ther field to have a constant norm
\begin{eqnarray}
u^a u_a = v^2,
\end{eqnarray}
where $\bar{\lambda}$ acts as a Lagrange multiplier and we take $v^2 > 0$. In our unit $u^a$ has dimension of mass$^{3/2}$. The sum $\mathcal{L}_i$ in (\ref{aether_action}) represent various interaction terms which couple the {\ae}ther field to matter fields that we will discuss later in this section.  If we neglect the interaction terms for the moment, the equations of motion for the {\ae}ther field $u^a$ can be written as
\begin{eqnarray}
\nabla_a V^{ab}+v^{-2}u^b u_c\nabla_dV^{cd} = 0.
\label{u_eom}
\end{eqnarray}
Any solutions for which $V_{ab} = 0$ will solve the equation of motion (\ref{u_eom}).  In order to preserve Lorentz invariance in the 4-dimensional non-compact space, we choose the background solution such that the {\ae}ther is a space-like vector field which has non-vanishing components along the extra fifth dimension, 
\begin{eqnarray}
u^a = (0, 0, 0, 0, v).
\label{flatspace}
\end{eqnarray}
It is important to note that there is a subtle stability issue here. Although our aim is to investigate the role of the {\ae}ther field on the stability of the extra dimension, the model of space-like {\ae}ther field with Maxwell-type kinetic term itself is {\it unstable} \cite{Carroll08}.  However, for our purpose, we can consider it as {\it a toy model} and assume that there is some mechanism which stabilizes the {\ae}ther field.

The energy-momentum tensor of the {\ae}ther field $T_{ab}|_{u}$ takes the following form
\begin{eqnarray}
T_{ab}|_{u}=V_{ac}V^{c}_b-
\frac{1}{4}V_{cd}V^{cd}g_{ab}+v^{-2}u_au_bu_c\nabla_dV^{cd}.
\label{T_u}
\end{eqnarray}
Note that properties of the {\ae}ther field depend crucially on spacetime geometry. The flat space background solution in equation (\ref{flatspace}) gives $T_{ab}|_u = 0$.  However, in curved spacetime, the {\ae}ther field can give rise to non-vanishing energy momentum tensor for example a time-like {\ae}ther field can produce energy density \cite{Carroll_Lim} while a space-like {\ae}ther gives the stress components \cite{Ackerman}.  The case of an {\ae}ther field oriented along the compact extra dimension was investigated in \cite{Carroll_Tam}.  It was shown that such {\ae}ther configuration can also give rise to non-vanishing energy momentum tensor.  However, $T_{ab}|_u$ vanishes when the extra dimension is stabilized.  We will review this result in the next section.

We now consider the effect of the interaction term $\sum_i \mathcal{L}_i$ in (\ref{aether_action}) which in general can include the terms corresponding to the {\ae}ther field  coupled to scalars, fermions and gravity.  However, we will consider stabilization mechanism of the extra dimension involving Casimir energy of gravitons, bulk scalars and bulk fermions.  We will ignore the bulk vector terms. Let us begin with the effect of the interaction of the {\ae}ther with a real massive scalar field $\phi$.  The Lagrangian for the scalar field with the minimal coupling term is
\begin{eqnarray}
\mathcal{L}_\phi=
-\frac{1}{2}(\partial_a\phi)^2-\frac{1}{2}m^2\phi^2-\frac{1}{2\mu_\phi^2}u^au^b\partial_a\phi\partial_b\phi,
\end{eqnarray}
where $\mu_\phi$ is the coupling parameters with dimension of mass$^{3/2}$. The corresponding equation of motion for the scalar field takes the form \cite{Carroll_Tam}\begin{eqnarray}
\partial_a\partial^a\phi -
m^2\phi=-\mu^{-2}_\phi\partial_a(u^au^b\partial_b\phi).
\end{eqnarray}
Expanding the scalar field in Fourier modes $\phi\propto e^{ik_ax^a}$, we obtain the modified dispersion relation,
\begin{eqnarray}
-k^\mu k_\mu = m^2 + (1+\alpha_\phi^2)k^2_5,\label{Sdis}
\end{eqnarray}
where the dimensionless parameter $\alpha_\phi=v/\mu_\phi$ is the ratio of the aether vev to the coupling $\mu_{\phi}$.  Next we consider
the fermion terms. The Lagrangian for fermionic field with the minimal coupling term can be written as \cite{Carroll_Tam}
\begin{eqnarray}
\mathcal{L}_\psi=i\bar{\psi}\gamma^a\partial_a\psi-
m\bar{\psi}\psi-\frac{i}{\mu_\psi^2}u^au^b
\bar{\psi}\gamma_a\partial_b\psi,
\end{eqnarray}
where $\mu_{\phi}$ is the fermionic coupling constant with the unit of mass$^{3/2}$.  In the same spirit as in the scalar field case, the corresponding modification of the dispersion relation for the fermionic case can be written as
\begin{eqnarray}
-k^\mu k_\mu = m^2 + (1+\alpha_\psi^2)^2k^2_5,
\label{Fdis}
\end{eqnarray}
where the dimensionless parameter $\alpha_\psi=v/\mu_\psi$. The form of this equation is different from the analogous equation in the bosonic case: i.e. the second term on the right-handed side increases by $\alpha_{\psi}^4$ instead of $\alpha_{\psi}^2$. Finally, we consider the {\ae}ther field which couples non-minimally to gravity.  This can be described by the action \cite{Carroll_Tam}
\begin{eqnarray}
S_{GC}= \int d^5x \sqrt{-g}\Big( \frac{M^3_{*}}{16\pi}R +\alpha_g u^a
u^bR_{ab} \Big)\label{action1},
\end{eqnarray}
where $\alpha_g$ is the dimensionless graviton coupling constant and $M_{*}$ is the Planck mass in 5 dimensional space-time. By varying this action with respect to the metric tensor, we obtain the equation of motion $G_{ab}= 8\pi G T_{ab}|_{(GC)}$ with 
\begin{eqnarray}
T_{ab}|_{(GC)}=\alpha_g\Big( R_{cd}u^cu^dg_{ab}
+\nabla_c\nabla_a(u_bu^c)+\nabla_b\nabla_c(u_au^c)-\nabla_c\nabla_d(u^cu^d)g_{ab}-\nabla_c\nabla^c(u_au_b)\Big),
\label{T_CG}
\end{eqnarray}
where $G$ is the 5-dimensional gravitational constant. Let us consider small fluctuation of the metric 
\begin{equation}
g_{ab} = \eta_{ab} + h_{ab}.
\end{equation}
Following the explanation in \cite{Carroll_Tam}, the metric perturbation can be decomposed into
\begin{equation}
h_{\mu\nu} = \bar{h}_{\mu\nu} + \bar{\Phi}\eta_{\mu\nu}, ~~h_{55} = \bar{\Psi},
\end{equation}
where $\eta^{\mu\nu}\bar{h}_{\mu\nu} = 0$, $\bar{h}_{\mu\nu}$ presents the propagating modes of the gravitational wave, $\bar{\Phi}$ denotes the Newtonian gravitational field and $\bar{\Psi}$ is a component associated with the radion field describing the modes of the extra dimension.  By setting $\bar{\Phi} = 0 = \bar{\Psi}$, and considering transverse waves, $\partial^{\lambda}\bar{h}_{\lambda\mu}=0$, the gravitational equation of motion becomes
\begin{equation}
-\frac{1}{2}\partial^{c}\partial_{c}\bar{h}_{\mu\nu} = 8\pi \frac{\alpha_g v^2}{M_{*}^3} \partial^2_5 \bar{h}_{\mu\nu}.
\end{equation}
Let us define $\tilde{\alpha}_g^2 = 16\pi\frac{\alpha_g v^2}{M_*^3}$.  The above equation gives the modified dispersion relation for graviton
\begin{equation}
- k^{\mu}k_{\mu} = \left( 1 + \tilde{\alpha}_g^2 \right)k_5^2.
\label{Gdis}
\end{equation} 

\section{Cosmological Dynamics and {\AE}ther Field}\label{Cosmology}
\subsection{Five-dimensional cosmological dynamics}
\label{5D}

In this section we consider cosmological dynamics of 5-dimensional spacetime by applying Einstein general relativity to the product space, between 4-dimensional FRW-type spacetime and a circle $S^1$.  We assume the cosmological ansatz
\begin{eqnarray}
ds^2 = -dt^2 + a(t)^2 dx^i dx^j\delta_{ij} + b(t)^2 dy^2,
\label{metric}
\end{eqnarray}
where $i,j = 1,2,3$,  $a(t)$ is the scale factor of non-compact 3-dimensional space, and $b(t)$ denotes the radius of the compact fifth direction. The coordinates on $S^1$ are $0 \leq y \leq 2\pi$.  For our metric (\ref{metric}), the background solution for the equation of motion (\ref{u_eom}) can be written as
\begin{eqnarray}
u^a = \Big(0, 0, 0, 0, \frac{v}{b(t)}\Big).
\label{u_bg}
\end{eqnarray}
Using this background solution, the energy momentum tensor associated to the {\ae}ther field defined in (\ref{T_u}) can be written as
\begin{equation}
T^0_{~0}|_u = -\frac{v^2}{2}H_b^2,~~
T^i_{~j}|_u = \frac{v^2}{2}H_b^2\;\delta^i_{~j},~~
T^5_{~5}|_u = - \frac{\ddot{b}}{b} + \frac{1}{2}H_b^2 - 3H_aH_b
\label{T_u_bg}
\end{equation}
We have defined the Hubble constants $H_a = \dot{a}/a$ and $H_b = \dot{b}/b$, where dotted quantities represent the corresponding time derivative. As we mentioned in the previous section, $T_{ab}|_u =0$ when the extra dimension is stabilized $\dot{b} =0$.  The fact that the {\ae}ther field does not contribute to the energy density at the stabilized point implies that the {\ae}ther field will not give any contribution to the effective potential of the radion. Hence other component such as Casimir energy is needed for stabilization of the extra dimension.  However, as we shall see later on, the {\ae}ther field can reduce the influence of the Casimir force.  This property is important for stabilization mechanism when non-relativistic matter is present.

Let us assume that the total energy-momentum tensor $T_{ab} |_{total}$ is decomposed into
\begin{equation}
T_{ab} |_{total} = T_{ab}|_{u} + T_{ab}|_{GC} + T_{ab} |_{\rho}.
\label{T_total}
\end{equation}
The contribution from non-minimally coupling to gravity $T_{ab} |_{GC}$ is defined in equation (\ref{T_CG}).  The component $T^a_b |_{\rho} = diag(-\rho,p_a,p_a,p_a,p_b)$ represents contribution from Casimir energy \cite{Greene}.  Casimir energy density $\rho$ plays the role of 5-dimensional cosmological constant. $p_a = -\rho$ and $p_b = -\frac{\partial (\rho 2 \pi b)}{\partial (2 \pi b)} = -\rho - b \partial_b \rho$ are the pressure density in non-compact and compact direction respectively.  By substituting $T_{ab}|_{total}$ into the Einstein field equation, we get the 5-dimensional cosmological equations of motion
\begin{eqnarray}
3H_a^2+ 3H_aH_b &=& 8\pi G(\rho+\frac{1}{2} v^2 H_b^2),\label{eom1}\\
3\frac{\ddot{a}}{a}- 3H_aH_b &=& - 8\pi G\Big\{\rho + p_b -  (1 - 2\alpha_g)v^2A\Big\},\label{eom2}\\
3\frac{\ddot{b}}{b} +9H_aH_b &=& 8\pi
G\Big\{\rho + 2p_b - 3p_a -2(1 - 2\alpha_g)v^2A\Big\},\label{eom3}
\end{eqnarray}
where $A=(\frac{\ddot{b}}{b} +3H_aH_b)$. 

\subsection{Dynamics in the radion picture}

Since we are interested in our observed universe, it is useful to analyze the cosmological dynamics by considering 4-dimensional effective field theory.  The equations of motion (\ref{eom1})-(\ref{eom3}) can be obtained by varying the 5-dimensional Einstein-Hilbert action
\begin{eqnarray}
S_{5D}= \int d^5x \sqrt{-g}\Big( \frac{M^3_{*}}{16\pi}R -\frac{1}{4}V_{a b}V^{a b} +\alpha_g u^a
u^bR_{ab} - V(b)\Big)
\label{action5D}.
\end{eqnarray}
$V(b)$ denotes the potential term in 5-dimensional spacetime.  Note that we omit the Lagrange multiplier term. For simplicity, we will set $\alpha_g = 0$ in this section and this will not affect our main results.  Let us start with KK-dimensional reduction of the above action from $5$ to $4$-dimensional spacetime.  Then, in order to make the resulting effective action in the canonical form, we apply Weyl rescaling $g_{\mu\nu E} = \Omega g_{\mu\nu}$ $(\mu,\nu = 1,\dots , 3)$ and define the new time variable $dt_E = \sqrt{\Omega} dt$, $a_E(t_E) = \sqrt{\Omega}a(t)$; $\Omega = 2 \pi b M_*^3/m_{pl}^2$. Note that $m_{pl}$ is the Planck mass in 4-dimensional spacetime defined via the relation $m_{pl}^2 = (2\pi b_{min})M_*^3$ where $b_{min}$ denotes the stabilized radius of extra dimension. Thus $\Omega = 1$ at $b = b_{min}$.  The effective action takes the form
\begin{eqnarray}
S_{4D}= \int d^4x \sqrt{-g_E}\Big\{ \frac{m^2_{pl}}{16\pi} R_E -\frac{1}{2}g_E^{\mu\nu}\nabla_{\mu}\Psi\nabla_{\nu}\Psi - \frac{1}{2}\frac{m_{pl}^2}{M_*^3b_{min}^2}e^{-2\frac{\sqrt{16\pi}}{\sqrt{3}m_{pl}}\Psi}V_{\mu}V^{\mu}  - U(\Psi)\Big\}
\label{action4D1},
\end{eqnarray}
where $U(\Psi) = 2\pi b\Omega^{-2}V(b)$ is the 4-dimensional effective potential.  Here we define the radion field $\Psi = \frac{m_{pl}}{\sqrt{16\pi}} \sqrt{3} ln(b/b_{min})$ and  $V_{\mu} = V_{\mu 5} = \nabla_{\mu}u_5$.  By using the background solution in (\ref{u_bg}), the above 4-dimensional action can be rewritten as
\begin{eqnarray}
S_{4D}= \int d^4x \sqrt{-g_E}\Big\{ \frac{m^2_{pl}}{16\pi} R_E -\frac{1}{2}(1+\alpha^2)g_E^{\mu\nu}\nabla_{\mu}\Psi\nabla_{\nu}\Psi  - U(\Psi)\Big\}
\label{action4D2},
\end{eqnarray}
where we define the dimensionless parameter $\alpha^2 = \frac{16\pi v^2}{3M_{*}^3}$. This action gives rise to the following set of equations:
\begin{eqnarray}
H_E^2 &=& \frac{8\pi}{3m_{pl}^2}\Big\{U(\Psi)+\frac{1}{2}(1+\alpha^2)\left(\frac{d\Psi}{dt_E}\right)^2\Big\},\label{reom1}\\
\frac{d^2\Psi}{dt_E^2} + 3 H_E \frac{d\Psi}{dt_E} &=& -\frac{1}{(1+\alpha^2)} \frac{\partial U}{\partial\Psi}.\label{reom2}
\end{eqnarray}
Note that $H_E = (da_E/dt_E)/a_E$ is the Hubble constant in the Einstein frame.  As we will explain later, the factor $1/(1+\alpha^2)$ in the right-handed side of equation (\ref{reom2}) weakens the effect of the potential gradient $-\partial U/\partial \Psi$ and it is crucial for stabilization mechanism of the radion field $\Psi$.  To make contact with previous section, we note that the energy-momentum tensor associated with 5-dimensional action in (\ref{action5D}) gives the relations
\begin{equation}
\rho = \frac{\Omega}{G m_{pl}^2}U \;,\; \; \; 2\rho + p_b = -\frac{\Omega}{Gm_{pl}^2}\left(b\partial_bU\right).
\end{equation}

\section{{\AE}ther Field and Casimir Energy}\label{Casimir}

We will start this section by reviewing the mathematical formulation to determine the Casimir energy for a scalar field, $\widehat{E}_{cas}$, and then investigating the effect of {\ae}ther coupling to the Casimir energy.  First, we consider Casimir energy of a non-interacting scalar field of mass, $m$, in $\mathbb{M}^{1+n}\times S^1$ spacetime by following \cite{Ambjorn, Elizalde1}. We keep the number of non-compact spatial directions to be $n$ for the moment and will set $n=3$ at the end of our calculation. This scalar field obeys the free Klein-Gordon equation,
\begin{eqnarray}
(\partial_a\partial^a - m^2)\phi=0.
\end{eqnarray}
The scalar field satisfies the periodic boundary condition in the compact direction, $\phi(y = 0)=\phi(y = 2\pi)$. Its associated dispersion relation can be written as
\begin{eqnarray}
-k^\mu k_\mu = m^2+\frac{\tilde{n}^{2}}{b^2},
\label{normal_dp}
\end{eqnarray}
where, $\tilde{n}\in\mathbb{Z}$ is the momentum number in the compact direction. Then, the total vacuum energy contributing to Casimir energy can be written as
\begin{eqnarray}
\widehat{E}_{cas} &=& \frac{1}{2}\left(\frac{L}{2 \pi}\right)^{n} \int d^{n}k
\sum_{\tilde{n}}\sqrt{ k^2+ m^2+ \frac{\tilde{n}^2}{b^2}},
\end{eqnarray}
where $V_{n}=L^{n}$ is the spatial volume of non-compact spacetime.  Using the fact that $\int f(k)d^n k = 2\pi^{n/2}/\Gamma(n/2)\int k^{n-1}f(k)dk$, we obtain
\begin{eqnarray}
\widehat{E}_{cas} &=& \frac{1}{2}\left(\frac{L}{2 \pi}\right)^{n}
\frac{2\pi^{n/2}}{\Gamma(n/2)} \int k^{n-1} dk \sum_{\tilde{n}}\sqrt{ k^2+
m^2+ \frac{\tilde{n}^2}{b^2}},\\
&=& \frac{1}{2}\Big(\frac{2}{L}\Big)^{2s+1}\frac{\Gamma(s)}{\Gamma(-1/2)}b^{2s}\pi^{(2s+1)/2}
\sum_{\tilde{n}}\Big((bm)^2+ \tilde{n}^2\Big)^{-s},
\end{eqnarray}
where we define $s= -(n+1)/2$. Let us consider the massless case, $m=0$.  By using the zeta function regularization procedure, the Casimir energy density per one bosonic degree of freedom for massless scalar field can be written as
\begin{eqnarray}
\widehat{\rho}_{cas}^{massless}=\frac{\widehat{E}_{cas}}{V_n 2\pi b}
=\frac{\Gamma(-2s+1)}{\Gamma(-1/2)}2^{2s}b^{2s-1}\pi^{3s-1}\zeta(-2s+1),
\end{eqnarray}
where $\zeta$ denotes the zeta function and we take $2 \pi b$ to be the volume of compact dimension. For the massive case, we apply the Chowla-Selberg zeta function \cite{Elizalde1} in our  regularization procedure and obtain the Casimir energy density per one degree of freedom for the massive scalar field:
\begin{eqnarray}
\widehat{\rho}_{cas}^{massive} = -2(2\pi
b)^{2s-1}(mb)^{(1-2s)/2}\sum_{n=1}^{\infty}n^{(2s-1)/2}K_{(1-2s)/2}(2\pi
bmn),
\end{eqnarray}
where $K_\nu(x)$ is the modified Bessel function. The fermionic degrees of freedom will contribute to the Casimir energy density with the same expression except for an extra minus sign.

Let us consider the case that a scalar field couples to an {\ae}ther field with a coupling constant $\alpha_{\phi}$.  In the previous section, we showed that interaction with the {\ae}ther field transforms the usual dispersion relation (\ref{normal_dp}) into its modified version (\ref{Sdis}). Accordingly, the Casimir energy will be written as
\begin{eqnarray}
E_{cas}(\alpha_{\phi}) &=& \frac{1}{2}\left(\frac{L}{2 \pi}\right)^{n} \int d^{n}k
\sum_{\tilde{n}}\sqrt{ k^2+ m^2+ (1+\alpha_\phi^2)\frac{\tilde{n}^2}{b^2}}, \nonumber\\
&=& \frac{1}{2}\left(\frac{L}{2 \pi}\right)^{n} \frac{2\pi^{n/2}}{\Gamma(n/2)}
\int k^{n-1} dk  \sum_{\tilde{n}}\sqrt{ k^2+ m^2+
(1+\alpha_\phi^2)\frac{\tilde{n}^2}{b^2}}, \nonumber\\
&=& (1+\alpha_\phi^2)^{(n+1)/2}\frac{1}{2}(\frac{L}{2 \pi})^{n}
\frac{2\pi^{n/2}}{\Gamma(n/2)} \int k'^{n-1} dk' \sum_{\tilde{n}}\sqrt{
k'^2+ m'^2+ \frac{\tilde{n}^2}{b^2}},
\end{eqnarray}
where we rescale $k$ and $m$ in such a way that $k^2=(1+\alpha_\phi^2)k'^2$ and $m^2=(1+\alpha_\phi^2)m'^2$. By comparing $E_{cas}(\alpha_{\phi})$ with the non-interacting Casimir energy $\widehat{E}_{cas}$, we see that the {\ae}ther coupling rescales the Casimir energy and scalar mass by factors $(1+\alpha_\phi^2)^{(n+1)/2}$ and $(1+\alpha_\phi^2)^{-1/2}$ respectively. Thus, we can immediately write down the Casimir energy density per one bosonic degrees of freedom with the {\ae}ther coupling $\alpha_\phi$ as
\begin{eqnarray}
\rho_{boson}^{massless}(\alpha_{\phi})&=&
\frac{\Gamma(-2s+1)}{\Gamma(-1/2)}\frac{2^{2s}b^{2s-1}\pi^{3s-1}}{(1+\alpha_\phi^2)^{s}}\zeta(-2s+1),\label{rho_boson1}\\
\rho_{boson}^{massive}(\alpha_{\phi}) &=& -\frac{2(2\pi
b)^{2s-1}}{(1+\alpha_\phi^2)^{s}}\Big(\frac{mb}{\sqrt{1+\alpha_\phi^2}}\Big)^{\frac{(1-2s)}{2}}
\sum_{\tilde{n}=1}^{\infty}\tilde{n}^{(2s-1)/2}K_{(1-2s)/2} \Big(\frac{2\pi
mb\tilde{n}}{\sqrt{1+\alpha_\phi^2}}\Big),
\label{rho_boson2}
\end{eqnarray}
for contributions from massless and massive scalar fields respectively.  Other bosonic degrees of freedom contribute to the total Casimir energy in a similar way but with different coupling constants and masses.  For example, the graviton in ($n+2$)-dimensional spacetime has $\frac{1}{2}(n+2)(n-1)$ bosonic degrees of freedom with the {\ae}ther coupling $\alpha_g$ and mass $m=0$. Using the modified dispersion relation for graviton (\ref{Gdis}), we can show that the graviton contributes $\frac{(n+2)(n-1)}{2}\rho_{boson}^{massless}(\tilde{\alpha}_{g})$ to the total Casimir energy density.  Recall that, apart from modification of graviton's Casimir energy, non-minimal coupling of   the {\ae}ther field to gravity can also affect the dynamical evolution through the energy momentum tensor $T_{ab}|_{(GC)}$. 

For fermion case, we use the modified dispersion relation for the fermionic field in (\ref{Fdis}).  We can show that the associated Casmir energy density per one fermionic degree of freedom, for both massless and massive case, is in the similar form of those from bosinic degrees of freedom with the over all minus sign and $(1+\alpha_\phi^2)\rightarrow(1+\alpha_\psi^2)^2$.  The Casimir energy densities for one degree of freedom of massless and massive fermion can be written respectively as
\begin{eqnarray}
\rho_{fermion}^{massless}(\alpha_{\psi})&=&-
\frac{\Gamma(-2s+1)}{\Gamma(-1/2)}\frac{2^{2s}b^{2s-1}\pi^{3s-1}}{(1+\alpha_\psi^2)^{2s}}\zeta(-2s+1),\label{rho_fermion1}\\
\rho_{fermion}^{massive}(\alpha_{\psi}) &=& \frac{2(2\pi
b)^{2s-1}}{(1+\alpha_\psi^2)^{2s}}\Big(\frac{mb}{1+\alpha_\psi^2}\Big)^{(1-2s)/2}
\sum_{n=1}^{\infty}n^{(2s-1)/2}K_{(1-2s)/2} \Big(\frac{2\pi
mbn}{1+\alpha_\psi^2}\Big).\label{rho_fermion2}
\end{eqnarray}
The total Casimir energy density can be written in terms of sum over all degrees of freedom: 
\begin{equation}
\rho = N_b \rho_{boson}^{massless}(\tilde{\alpha}_g) + N_f \rho_{fermion}^{massless}(\alpha_{\psi}) + \tilde{N}_b \rho_{boson}^{massive}(\alpha_{\phi}) + \tilde{N}_f \rho_{fermion}^{massive}(\alpha_{\psi}),
\label{sum_casimir}
\end{equation}
where $N_b$ ($N_f$) and $\tilde{N}_b$ ($\tilde{N}_f$) are the numbers of bosonic (fermionic) degrees of freedom for massless and massive fields respectively.  The nature of the total Casimir energy density depends on the relative magnitude of $N_b$, $N_f$, $\tilde{N}_b$ and $\tilde{N}_f$. 

In our model, $N_b \geqslant 5$ since, at least, the graviton is always present and it has five physical degrees of freedom in five-dimensional spacetime ($n =3$).  The compact fifth direction would not be stable if there is only the graviton field in the bulk, i.e. it will collapse to Planck size, due to the negative Casimir energy associated with quantum fluctuations of the gravitation fields. Therefore, it is natural to add more positive contribution to the Casimir energy by assuming that there are fermions in the bulk.  However, we cannot create the minimum of $\rho$ by including only the massless fermionic fields i.e. the Casimir force is attractive for $N_f < N_b$ and becomes repulsive when $N_f > N_b$. Hence $N_f = N_b$ does not give us any stable fixed point.  We should expect a minimum to be produced if we include massive fermionic degree of freedom \cite{Rubin}.  This can be explained qualitatively as the following.  Let us consider Casimir energy of a fermion with mass $M$.  For the region where $b \ll 1/M$, the vacuum energy should have the same form as in the massless case.  In particular, for $N_f > N_b$, the net Casimir force will be repulsive.  In the other region where $b \gg 1/M$, the contribution from the massive fermion mode is negligible compared to that of the graviton and the total Casimir force becomes attractive. Hence, there must be a stable fixed point between these two regions.  

In this paper, we consider a toy model where the particle spectrum in the bulk consists of a bulk graviton, a massless Dirac fermion, a massive Dirac fermion with mass $m_f$, and eight massive scalars with equal masses $m_s = \lambda m_f$.  Here $\lambda$ is the mass ratio. The presence of the massless fermion and the massive scalars is to ensure that the vacuum energy at the minimum has positive value, i.e. $\rho_{min} > 0$.  We summarize the particle content in the bulk in Table 1.  

\begin{table}[htdp]
\begin{center}\begin{tabular}{|c|c|c|c|}\hline 
Particles & Degrees of freedom & Mass & Coupling constant\\\hline 
a bulk graviton field & $N_b = 5$ &  $0$ & $\tilde{\alpha}_g = (16\pi\frac{\alpha_g v^2}{M_*^3})^{1/2} $\\
a massless bulk fermion field & $N_f = 8$ & $0$ & $\alpha_{\psi}$ \\
a massive bulk fermion field & $\tilde{N}_f = 8$ & $m_f$ & $\alpha_{\psi}$ \\
8 massive bulk scalar fields & $\tilde{N}_b = 8$ & $m_s = \lambda m_f$ & $\alpha_{\phi}$ \\\hline 
\end{tabular} 
\caption{The particle spectrum in the bulk, their degrees of freedom, their mass and the coupling constants characterizing their interaction with the {\ae}ther field. For simplicity, we assume the universal fermionic coupling $\alpha_{\psi} $ for both massless and massive Dirac fermion. }
\end{center}\label{table1}\end{table}

There is no unique choice of bulk particle spectrum for this purpose, the other combinations of bulk fields, for example in Ref.~\cite{Greene}, can probably create minimum for the vacuum energy.  Our particular choice is convenient for investigating the effect of the {\ae}ther-matter interactions on the Casimir energy of the bulk fields. Note that we do not attempt to justify the existence of these bulk fields phenomenologically because we want to demonstrate that stability of the extra dimension could be achieved, if these particles are present in the bulk.  The Dirac fermions in five dimensional spacetime have eight physical degrees of freedom.  For more realistic models which have chiral fermion on the brane, we can impose the orbifold reflection symmetry: $y \rightarrow -y$ on the compact direction.  However, this will not affect our main results on stabilization of the extra dimension.  We will ignore this issue for simplicity.  

\section{Effects of The {\AE}ther Field on Stabilization of The Extra Dimension}\label{stability}

\subsection{Stabilization in vacuum dominated universe}

\begin{figure}[htp]
\centering
\includegraphics[width=1.0\textwidth]{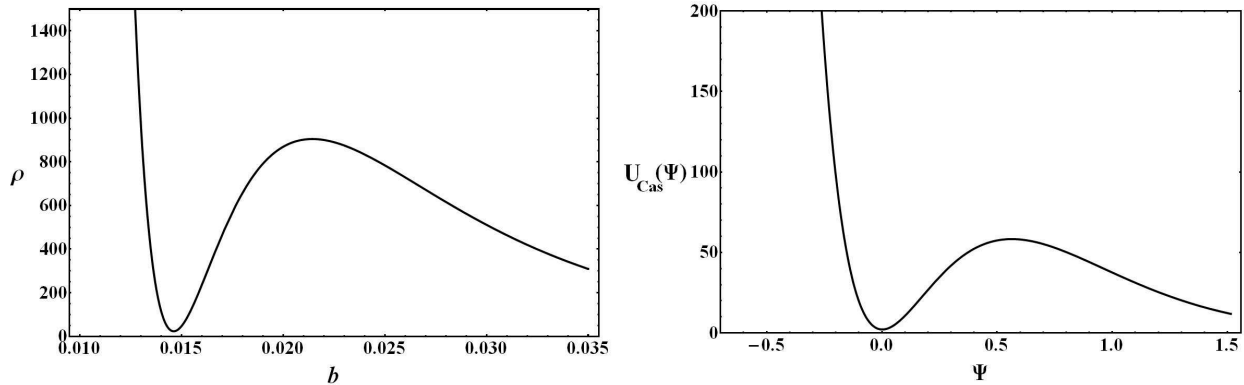}
\caption[$\rho_{cas}$]{Left: Casimir enery density $\rho$ is presented in the y-axis in units of $(m_f/40)^5$. The x-axis is the radius of the fifth dimension $b$ in units of $40/m_f$.  Right: The 4-dimensional effective Casimir potential $U_{Cas} =2\pi (b_{min}^2/b(\Psi))\rho(\Psi) $ in unit of $(m_f/40)^4$ is presented in the y-axis.  The x-axis denotes $\Psi$ in the unit of $m_{pl}$.  Here we set $\alpha=\alpha_{\phi}=\alpha_{\psi}=\alpha_g=0$.}\label{potential1}
\end{figure}

\begin{figure}[htp]
\centering
\includegraphics[width=1.0\textwidth]{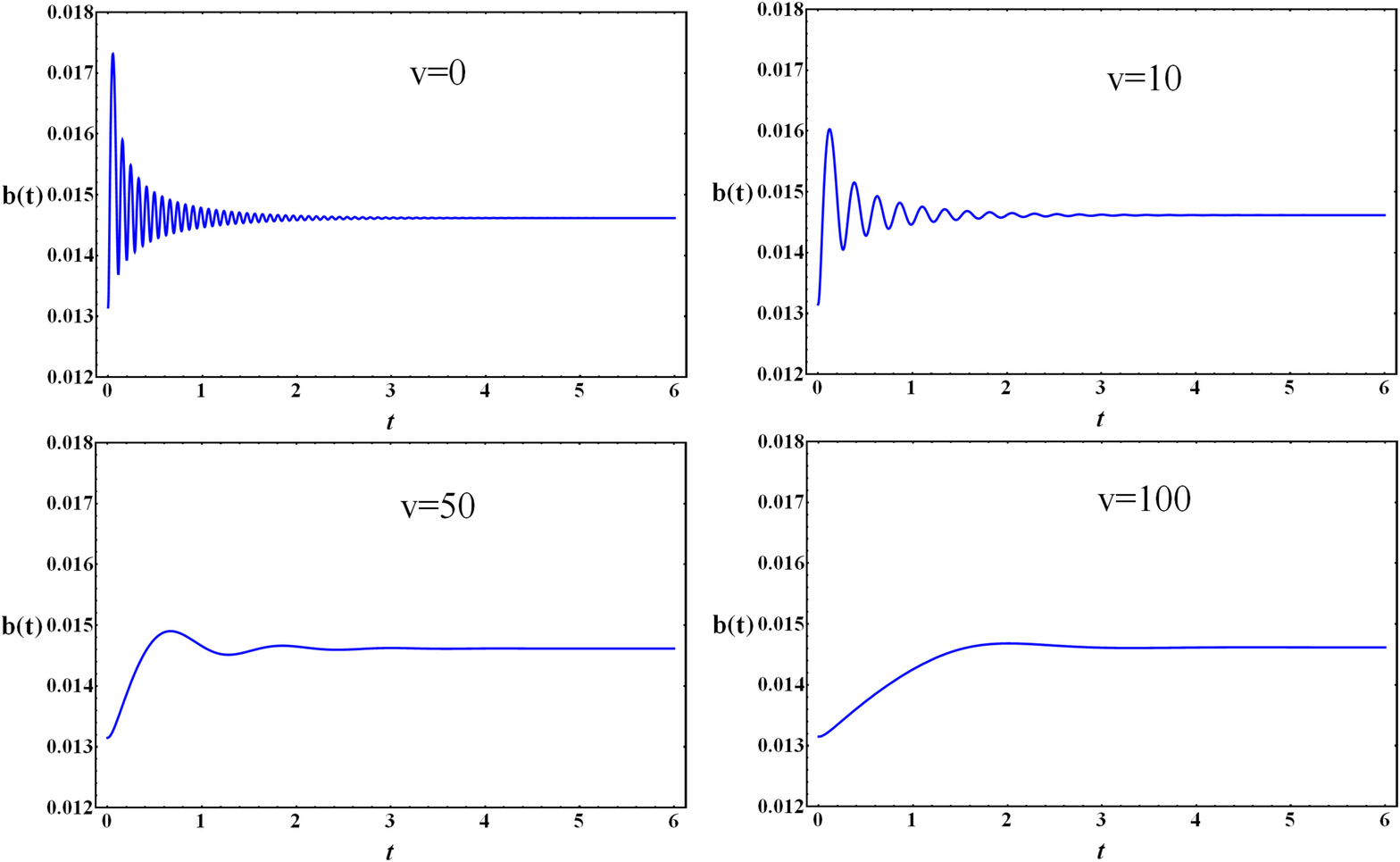}
\caption[$\rho_{cas}$]{The dynamics of the scale factor $b(t)$ for the compact direction as a function of time with different values of parameter $v$.  In the absence of {\ae}ther field $v=0$, $b(t)$ shows oscillation behavior around its critical value $b_{min}$ before stabilizing at this value.  Non-vanish value of $v$ reduces the influence of Casimir force.  As the value of $v$ increases, the oscillation frequency and amplitude decrease. If the vev of the {\ae}ther field is large enough, for example $v = 100$, oscillation behavior disappears.  The extra dimension evolves smoothly to its stable fixed point.  The time variable $t$ is presented in the unit of Hubble time $t_H$.  The time for stabilization to occur is around $\sim 6t_H$.  The condition for stabilization of $b$ is $\frac{\delta b}{b} \leqslant 10^{-5}$.}
\label{vacuum_dynamics}
\end{figure}

We first consider the universe where there is no non-relativistic matter and the Casimir energy density is the dominant contribution.  Let us start with the case where there is no interaction between {\ae}ther field and bulk matter i.e.  setting $\alpha_{\phi} = \alpha_{\psi} = \alpha_{g}= 0$.  The plot of the total Casimir energy density in 5-dimensional spacetime $\rho$ using the expression in equation~(\ref{sum_casimir}) and particle spectrum in Table 1, with the mass ratio $\lambda=0.516$, and its corresponding  4-dimensional effective $U_{Cas}=2\pi (b_{min}^2/b(\Psi))\rho(\Psi)$ is given in Figure~\ref{potential1}. The local minimum of Casimir energy density, $\rho_{min} = 23.4316 (m_f/40)^5$ is located at $b_{min} = 0.01461 (40/m_f)$.  In order to get the positive minimum and the radion can stabilize, we must choose the value of $\lambda$ from a very narrow range, $0.516\leq\lambda\leq0.527$.  By solving the 5-dimensional equations of motion (\ref{eom1})-(\ref{eom3}) numerically, we can demonstrate that the extra dimension is stabilized at  the radius $b_{min}$ as shown in Figure \ref{vacuum_dynamics}.  Notice that we set the expansion time scale to be in the unit of Hubble time $t_H = H_{a0}^{-1} = \sqrt{3 m_{pl}^2 / 8\pi \rho_c} \approx 10^{10}$ years.  The critical density $\rho_c$ can be written in terms of the minimum value of Casimir energy density $\rho_{min}$ as $\rho_c = (1+0.24/0.76)(2\pi b_{min})\rho_{min}$.  

By comparing 4-dimensional effective Casimir energy density $\rho^{(4)}_{min} = (2\pi b_{min})\rho_{min}$ with the observed value of energy density for dark energy, $\rho^{(4)}_{obs} \approx (2.3 \times 10^{-3}\;eV)^4$, we get $m_f \approx 4.18 \times 10^{-2}\; eV$.  Then, the radius of extra dimension $b_{min} \sim 13.96 \; eV^{-1} \sim 2.75 \times 10^{-6} \;m$.  This leads to the quantum gravity scale in the bulk, $M_* \approx 1.19 \times 10^{9}\;GeV$. Note that we do not attempt to address the mass hierarchy problem in this paper.  In order to compare with the ADD brane world scenario, it is better to be demonstrated with 6-dimensional models as shown in \cite{Greene}. Since our aim is to study the role of {\ae}ther fields on stabilization of the extra dimension, the 5-dimensional model is good enough for our purpose.  We will leave the mass hierarchy problem for future works.

The role of {\ae}ther field on dynamical evolution of the extra dimension is also illustrated in Figure \ref{vacuum_dynamics}.  We can see that, as the value of $v$ increase, the moduli field $b$ feel less potential gradient.  Its oscillation frequency and amplitude decrease.  In our caculation $v$ is in the unit of $(m_{f}/40)^{5/2}t_{H}$.  From the previous paragraph $m_f = 4.18 \times 10^{-2}\;eV$, this gives 1 unit of $v$ is equivalent to $(1.02\times 10^{8} \;GeV)^{3/2} \approx (0.09 M_*)^{3/2}$.  At very high $v$, for example $v = 100$ or approximately $ \approx (9 M_*)^{3/2}$, the period of oscillation is so long that $b$ reaches equilibrium before showing any oscillating behavior.  The scale factor $b$ rolls slowly to its stable fixed point.

\begin{figure}[htp]
\centering
\includegraphics[width=0.6\textwidth]{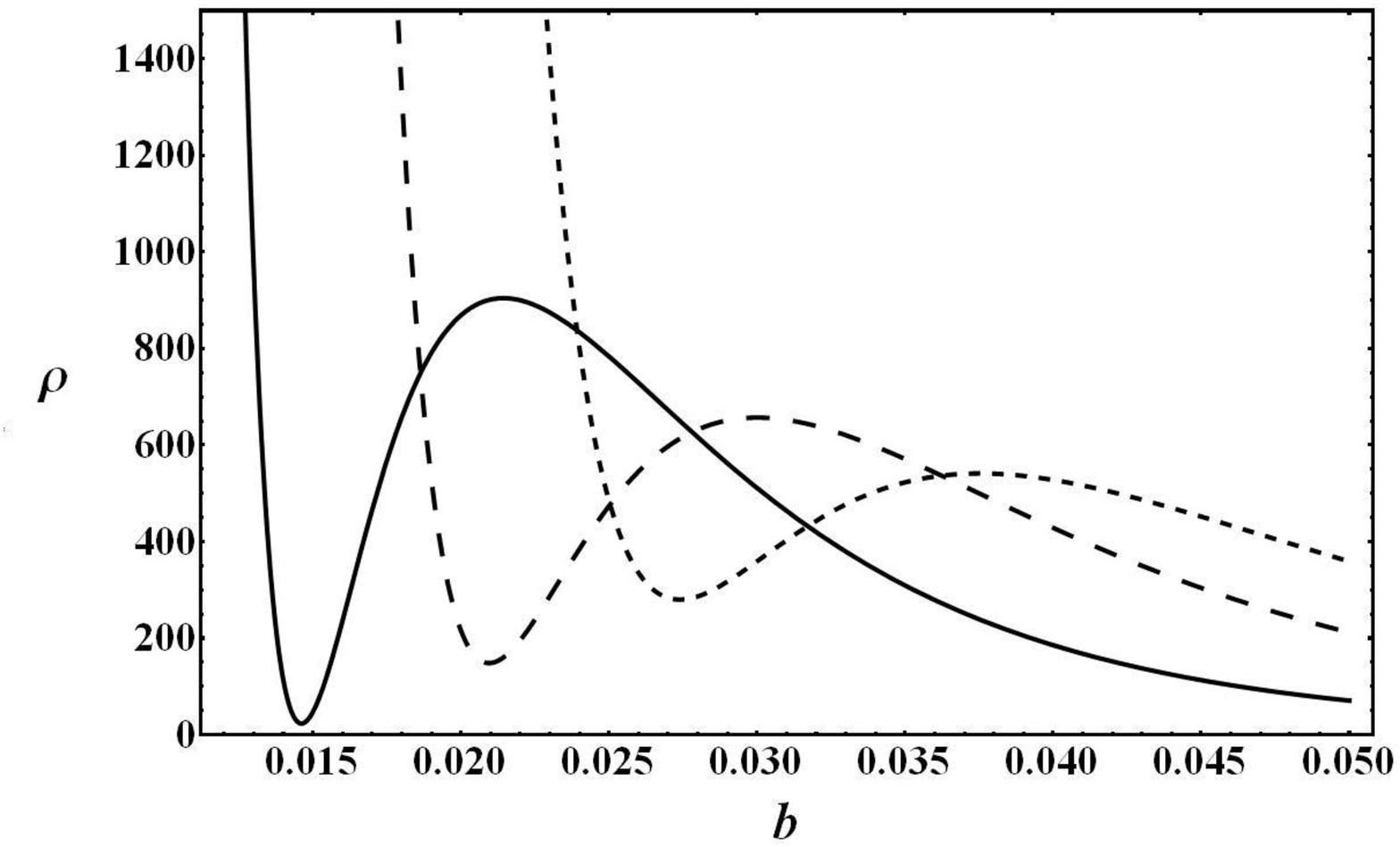}
\caption[$constraint1$]{Interactions between bosons/fermions and {\ae}ther field can affect the Casimir energy.  In this figure, we fix the mass ratio $\lambda = 0.516$.  The Casimir enery density $\rho$ is presented in the y-axis in units of $(m_f/40)^5$. The solid, long dashed and short dashed line denote the Casimir energy density when the coupling constant $(\alpha_\phi, \alpha_\psi) = (0.0, 0.0), (1.0, 0.644),$ and $(1.5, 0.897)$ respectively.  The value of $b_{min}$ and $\rho_{min}$ increase as we increase the value of the coupling constants. The shape of the potential well gets shallower as the coupling increases. We set $\tilde{\alpha}_g = \alpha_\phi$ for simplicity.}
\label{rhoap}
\end{figure}

Let us consider the situation where {\ae}ther field couples to the bulk matters.  As we discussed earlier, interactions with the {\ae}ther field reduce the effective mass of the bulk fields.  For example, the effective mass for scalar field of mass $m_s$ would be
\begin{eqnarray}
m^2_{s(eff)}=m^2_s(1+\alpha_\phi^2)^{-1}\label{masseff}.
\end{eqnarray}
This will alter the shape of the potential as we demonstrate in Figure \ref{rhoap}.  Interestingly, there is an advantage of coupling the bulk fields with the {\ae}ther field.  It seems that we get the wider range of parameter space for the mass ratio $\lambda$ that allows positive minima $\rho_{min} > 0$, i.e $0.05\lesssim\lambda\lesssim0.80$.

\subsection{Stabilization in the universe with non-relativistic matter}

In this section we consider the role of {\ae}ther field on stabilizing mechanism of the extra dimension in the more realistic model of our universe i.e. a model containing non-relativistic matter. Let us fist demonstrate the destabilizing effect due to non-relativistic matter by following Greene and Levin in Ref \cite{Greene}.  We assume that there is matter living in the bulk.  This can be done by adding a 5-dimensional matter term into the 5-dimensional action in equation (\ref{action5D})
\begin{equation}
S =  S_{5D} + \int d^{5}x \sqrt{-g}  \mathcal{L}_{matter} .
\end{equation} 
This is equivalent to adding the matter energy density $\rho_m\propto 1/2\pi a^3b$ into the 5-dimensional cosmological equations of motion (\ref{eom1})-(\ref{eom3}).  The matter energy density $\rho_m$ includes contribution from baryonic matter and cold dark matter. By comparing with the observational data and supposing that all dark matter is cold, the matter density today $\rho_{m0}$ is roughly 26\% of the total energy density of the universe.  The Casimir energy density will be responsible for the other 74\% of the total energy density today in the form of dark energy, $(\rho_{\Lambda0})$.  Thus, we have the relation, $\rho_{m0} = (2.6/7.4)\rho_{\Lambda0}$.  Note that the energy density of dark energy in our 4-dimensional observed universe today can be written in terms of the minimum of 5-dimensional Casimir energy density and the stabilized radius of the extra dimension as $\rho_{\Lambda0} = \rho_{min} (2\pi b_{min})= (2.3\times 10^{-3} eV)^{4}$.  By using $(a_0/a) = 1+z$, $a_0$ is the scale factor today and $z$ is the red-shift, we get
\begin{equation}
\rho_m = \frac{2.6}{7.4}\rho_{min}\left(\frac{b_{min}}{b}\right)(1+z)^3.
\end{equation}
In this case, the 5-dimensional equations of motion (\ref{eom1})-(\ref{eom3}) become 
\begin{eqnarray}
3H_a^2+ 3H_aH_b &=& 8\pi G(\rho + \rho_m + \frac{1}{2} v^2 H_b^2),\label{meom1}\\
3\frac{\ddot{a}}{a}- 3H_aH_b &=& - 8\pi G\Big\{\rho + \rho_m + p_b - (1-2\alpha_g)v^2A\Big\},\label{meom2}\\
3\frac{\ddot{b}}{b} +9H_aH_b &=& 8\pi
G\Big\{\rho + \rho_m + 2p_b - 3p_a -2(1-2\alpha_g)v^2A\Big\}.\label{meom3}
\end{eqnarray}
From equation (\ref{meom3}), the stability conditions ($\dot{H}_b =0$, $H_b =0$ and $A = 0$) require
\begin{equation}
p_b = -2 \rho - \frac{1}{2}\rho_m.
\label{pressure_constrain}
\end{equation}
By using equation (\ref{meom2}) and requiring that $\frac{\ddot{a}}{a} >0$ when $b$ is stabilized, we get the constraint $\rho_m < 2 \rho$.  This is the same constraint that we have for a $(3+1)$-dimensional vacuum dominated universe.  Thus, as pointed out in \cite{Greene}, this model describes the $(3+1)$-dimensional vacuum dominated universe when the extra dimension is stabilized.

By using the conservasion of the energy-momentum tensor in four and five dimensions, we can easily show that the radion field will be driven toward the minimum of the 4-dimensional effective potential
\begin{equation}
U_{eff} = U_{Cas} + \frac{m_{pl}^2G}{\Omega}\frac{\rho_m}{4} = U_{Cas} + \frac{\rho_m^{(4)}}{4}\left(\frac{b_{min}}{b}\right)^2,
\label{Umatter}
\end{equation}
where we define the 4-dimensional matter density $\rho_m^{(4)} = \rho_m (2\pi b)= \frac{2.6}{7.4}\rho_{min}\left(2 \pi b_{min}\right)(1+z)^3$ which is a function of $(1+z)^3$ and does not depend on the radius of the extra dimension $b$. 
\begin{figure}[htp]
\centering
\includegraphics[width=0.8\textwidth]{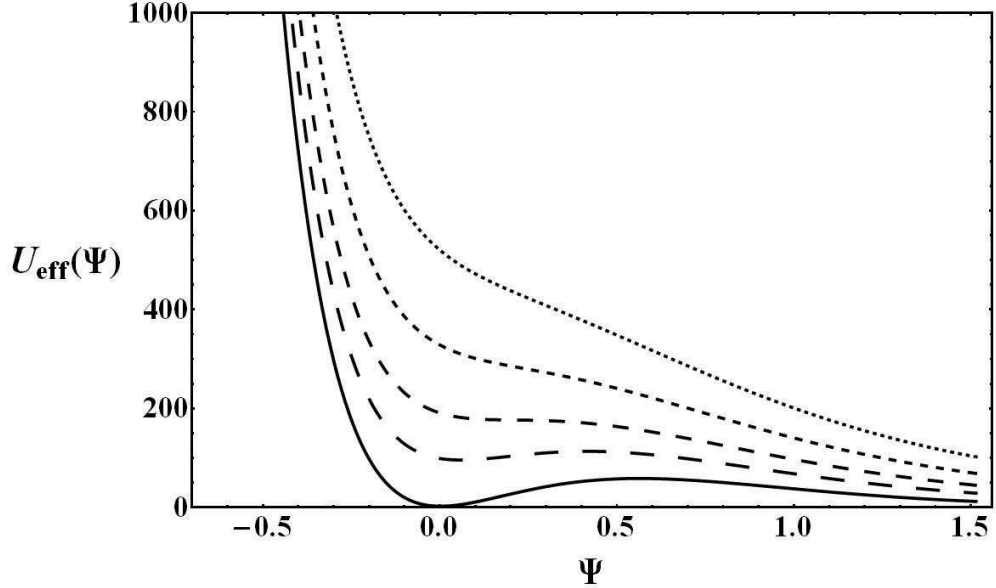}
\caption[$\rho_{cas}$]{This is the plot of the effective potential $U_{eff}(\Psi)$ at red shift $z= 0.0$, $ 7.0$, $9.0$, $11.0$, $13.0$ in the unit of $(m_f/40)^4$ and $\lambda = 0.516$. The local minimum of $U_{eff}(\Psi)$ no longer exists when the red shift is increased. }\label{f_umatt}
\end{figure}

 Numerical results for the effective potential $U_{eff}$ are illustrated in Figure \ref{f_umatt}.  Here we choose $\lambda = 0.516$, and ignore the interaction terms by setting $\alpha_{\phi}=\alpha_{\psi}=\alpha_g =0$.  At $z=0$, the presence of non-relativistic matter would lift up the minimum of $U_{eff}$ slightly.  However, at early time, high red-shift, the $1/b^{2}$-term in (\ref{Umatter}) becomes dominant and destroys the presence of the minimum. This effect will drive $b$ to expand even though there is a minimum today since the radion field $\Psi$ has already rolled pass the minimum and cannot get back to the stable point.  Notice that this effect is the same if matter is confined to the brane.  This can be shown by adding a 4-dimensional term into the 5-dimensional action (\ref{action5D})
  \begin{equation}
S =  S_{5D} + \int d^{4}x \sqrt{-g^{(4)}}  \mathcal{L}_{matter} .
\end{equation} 
There is no dimension reduction in the matter term but the conformal transformation rescale the matter energy density as $\rho_m^{(4)}/\Omega^2$.  Thus, the total effective potential is the same as in (\ref{Umatter}).

\begin{figure}[htp]
\centering
\includegraphics[width=1.0\textwidth]{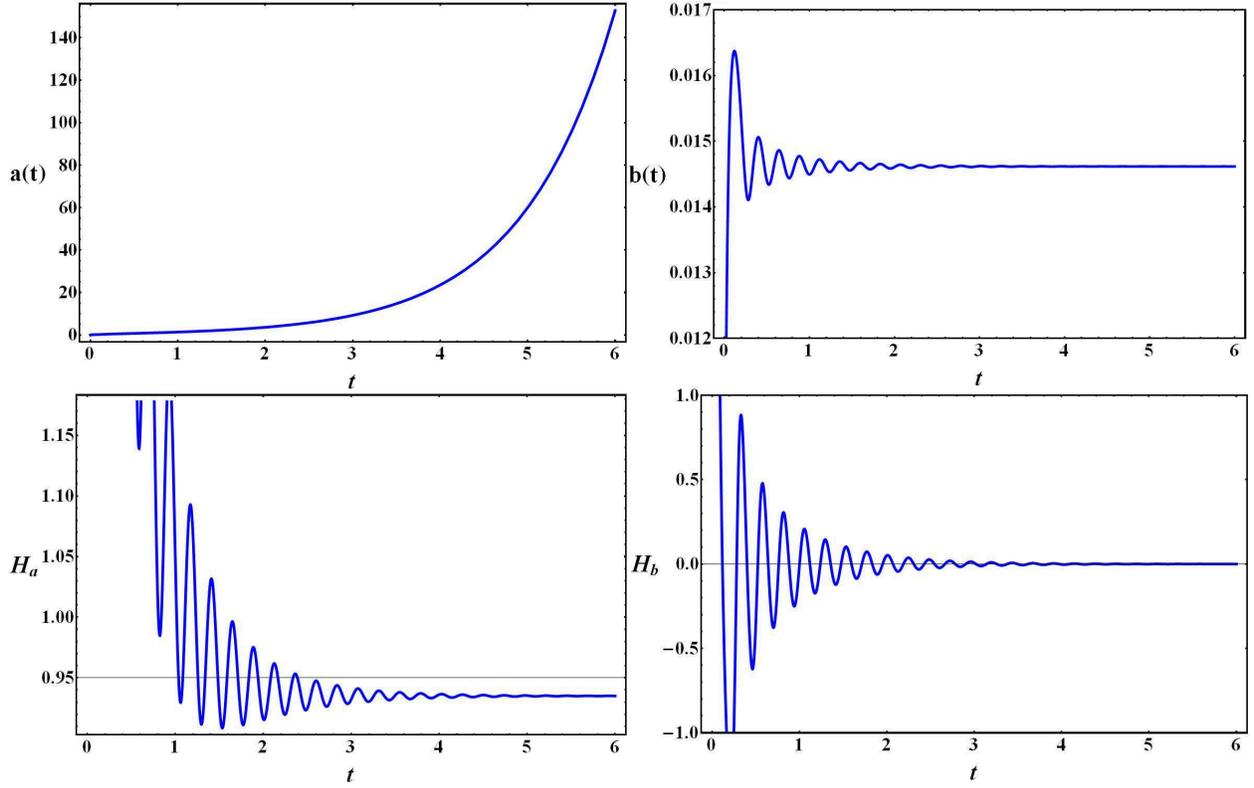}
\caption[$\rho_{cas}$]{These graphs illustrate cosmological dynamics of the universe which includes non-relativistic matter content and {\ae}ther field in the extra dimension.   Left: The scale factor $a$ (upper) and the Hubble constant for the non-compact dimensions $H_a$ (lower) as the function of time.  The fact that $H_a$ oscillates indicating the period of deceleration and acceleration before it settles down to the constant value and $a$ enters a de Sitter phase.  Right: The scale factor $b$ (upper) and the Hubble constant for the compact extra dimension $H_b$ (lower) as the function of time. $H_b$ oscillates between positive and negative region before settles down to zero.  The extra dimension is stabilized although non-relativistic matter is present.}\label{aether_stable}
\end{figure}

There is an interesting phenomena if we turn on the {\ae}ther field.  By carefully tuning its norm, we can show that $b$ can be stabilized even if there is non-relativistic matter in the universe.  This can be shown by solving equations of motion (\ref{meom1})-(\ref{meom3}) numerically.  The result is shown in Figure \ref{aether_stable} where we choose $\lambda = 0.516$, $m_f = 40$ and  $v =10$ (or equivalent to $v \approx  (0.9 M_*)^{3/2}$). We set the initial conditions such that $b(t=0) = 0.60 b_{min}$ and ignore the effects of interaction terms by setting $\alpha_{\phi}=\alpha_{\psi}=\alpha_g =0$.  From Figure \ref{f_umatt}, the minimum of $U_{eff}(\Psi)$ disappear when $z\gtrsim 11$.  So, in order to make sure that the minimum of $U_{eff}(\Psi)$ does not exist in the initial configuration, we choose the initial time to be at $z = 10,000$ which is approximately the time of matter-radiation equality ($t_{eq}$). Moreover since the universe also contains radiation component which is not considered in this model, it is reasonable to set $t_{initial} = t_{eq} = 0$.  However, if we ignore the radiation part, we can go back to the much earlier time and show that there are some values of $v$ that the extra dimension can be stabilized. 

The role of the {\ae}ther field can be explained qualitatively in the radion picture.  The {\ae}ther factor $1/(1+\alpha^2)$ on the right-handed side of equation (\ref{reom2}) reduces the influence of the potential gradient $-\partial U_{eff}/ \partial \Psi$.  As a consequence, it will slow down the oscillation frequency of $\Psi$ around the minimum of the potential  $U_{eff}(\Psi)$.  If this factor is big enough, $\Psi$ will move down the potential at very slow speed.  We can tune $v$ such that there is enough time for the universe to create the minimum of $U_{eff}(\Psi)$ before the radion rolls pass it.  By this mechanism, stability of the extra dimension can be restored.   

Let us compare the stabilization time $t_{stab}$ of the moduli field with the age of the universe.  The age of the universe in our model is
\begin{equation}
t_{age} = \frac{1}{H_{a0}}\int_0^{1}\frac{dx}{x\sqrt{\Omega_{Casimir}+\Omega_m x^3}}=\frac{1.5376}{H_{a0}} \approx 1.5376 t_H ,
\end{equation}
where we set $\Omega_{Casimir} = 0.76$ and $\Omega_m = 0.24$.  From Figure \ref{aether_stable}, the stabilization time $t_{stab} \approx 6 t_H$.  Then, $t_{stab} \approx 3.90 t_{age}$ is greater than the age of the universe.

\section{Conclusions and Discussions}\label{Conclusions}

In our 5-dimensional model, we have shown that the {\ae}ther field reduces the influence of the potential gradient and slows down the oscillation frequency of the compact extra dimension.  For vacuum dominated universe, the Casimir energy from the extra dimension acts as a stabilizing potential for the moduli field while driving accelerated expansion in the non-compact directions.  The {\ae}ther field will slow down the oscillation behavior of the moduli or even smooth it out.  For the universe which non-relativistic matter is present, this effect is proved to be crucial for stabilization of the extra dimension.  If the vev of {\ae}ther field is of the order of the 5-dimensional Planck mass, $v \sim O(M_*^{3/2})$, it can slow down the evolution of the moduli field such that there is enough time to create the minimum for the effective potential.  

In this paper we assume homogeneous and isotropic distribution of non-relativistic matter. However, local matter distributions might perturb the radion and knock it over the minimum, causing the (local) catastrophic expansion of the fifth dimension. In \cite{Greene}, it was also noted that the minimum of the potential well is generally not deep enough to prevent the quantum tunneling of the radion. At this stage, it is not clear whether these two difficulties can be solved by the new mechanism. These aspects of instability in the presence of the {\ae}ther are still open questions.
 
Note that the constancy of the 4-dimensional cosmological constant up to very early epoch of the universe will post strong constraint on the size of the extra dimension. The oscillation behavior of the moduli field may contradict with astronomical observations.  In order to construct a more realistic cosmological model of this scenario, the extra dimension should reach its stable fixed point before the present time i.e. $t_{stab} \lesssim t_{age}$ which require fine tuning of many parameters.  The new possible solution is that we assume very high value of $v$ so that the oscillation of $b$ has a very long period.  The moduli will evolve smoothly with no oscillating behavior.  We can choose the value of $v$ such that the size of the extra dimension changes so slowly and it cannot alter the results of the Big Bang model.

Another interesting idea is to imagine that the universe started with a very symmetric state which all spatial dimensions are compactified with the equal radius, for example, topologically an $4$-dimensional torus.  In general, the Casimir energies in this compact universe generate stabilized potential for the radius of all directions.  On the other hand, at the very early time, the energy density of matter and radiation will be the dominant contribution.  This will destabilize the moduli fields and all directions will become large.  However, if the Lorentz symmetry is spontaneously broken in some direction i.e. there is a non-vanishing {\ae}ther field pointing in the fifth direction, it will slow down the dynamics of moduli field associated to the broken direction.  The broken direction will be compactified at stabilized radius while the unbroken directions are allowed to expand.  This cosmological scenario may establish a connection between the dimensionality of spacetime and the violation of Lorentz symmetry.  We leave this issue for future investigation.

\section*{Acknowledgments}
\indent We would like to thank Piyabut Burikham,  Ahpisit Ungkitchanukit and Mark B. Wise for valuable discussions. A.C. is supported in part by the Thailand
Research Fund (TRF) and the Commission on Higher Education (CHE) under grant MRG5180225.

\end{document}